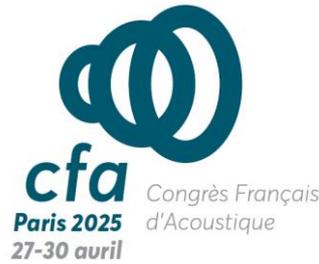

17e Congrès Français d'Acoustique 27-30 avril 2025, Paris

# Acoustic evaluation of a neural network dedicated to the detection of animal vocalisations

J. Rouch [a], M. Ducrettet [a,b,c], S. Haupert [c], R. Emonet [d] et F. Sèbe [a,e]

[a] ENES Bioacoustics Research Lab, CRNL, CNRS, INSERM, University of Saint-Etienne, Saint-Etienne, France

[b] BioPhonia, Sualello 20232, Oletta, France

[c] Institut Systématique Evolution Biodiversité (ISYEB), Muséum National d'Histoire Naturelle, CNRS, Sorbonne Université, EPHE, Université des Antilles, Paris, France

[d] Laboratoire Hubert Curien, Université de Saint-Etienne, CNRS, Institut d'Optique Graduate School, Inria, Saint-Etienne, France

[e] Office Français de la Biodiversité, Service Anthropisation & Fonctionnement des Ecosystèmes Terrestres, Direction de la Recherche et de l'Appui Scientifique, Gières, France

The accessibility of long-duration recorders, adapted to sometimes demanding field conditions, has enabled the deployment of extensive animal population monitoring campaigns through ecoacoustics. The effectiveness of automatic signal detection methods, increasingly based on neural approaches, is frequently evaluated solely through machine learning metrics, while acoustic analysis of performance remains rare. As part of the acoustic monitoring of Rock Ptarmigan populations, we propose here a simple method for acoustic analysis of the detection system's performance. The proposed measure is based on relating the signal-to-noise ratio of synthetic signals to their probability of detection. We show how this measure provides information about the system and allows optimisation of its training. We also show how it enables modelling of the detection distance, thus offering the possibility of evaluating its dynamics according to the sound environment and accessing an estimation of the spatial density of calls.

# 1   Introduction

The increasing use of ecoacoustics to monitor animal populations relies on the deployment of autonomous recorders[1]. The substantial volume of recordings often obtained, pushes towards the use of automatic detection approaches whose performance evaluation is based almost exclusively on metrics derived from the confusion matrix, such as precision and recall, which measure missed detections and false detections respectively. However, these metrics do not allow for fine analysis of the detector's ability to identify the target call, particularly its robustness to background noise.

In communication studies, this robustness is commonly evaluated by the average detection probability $p(snr)$, as a function of the signal-to-noise ratio $snr$[2]. For $L_U$ being the useful signal level in dB and $L_n$ the noise level in dB:

$$snr = L_U - L_n \qquad (1)$$

In audiology, for example, the $p(snr)$ curve allows estimation of a participant's level of hearing loss by comparing it to those of a normal-hearing population[3]. At low noise levels (high $snr$), noise does not affect detection and $p(snr)$ tends towards 1 for a high-performing system. Conversely, under excessive noise, the target signal becomes undetectable, and $p(snr)$ tends towards 0. The position, shape and width of the transition zone between these two extremes allow characterisation of the system's performance in noisy environments.

If the noise level at the receiver, the power of the source, and the propagation conditions are known, $p(snr)$ can be directly related to the maximum detection distance [4, 5] and therefore to the active area of the detector. However, in a population ecology study framework, without estimation of the active detection space and its variability, temporal comparisons within the same site and spatial comparisons between sites risk reflecting variations in detection areas rather than genuine ecological processes.

As part of the development of a neural network for monitoring populations of Rock Ptarmigan (*Lagopus muta*), a bio-indicator species of Arctic and mountain environments, we propose to integrate into its performance evaluation the estimation of its robustness as a function of the signal-to-noise ratio. The Materials and Methods section presents the neural network, its training, the evaluation of $p(snr)$ for different noise levels and the modelling of the active detection area. The Results detail the effect of data augmentation adjusted to $p(snr)$, and the influence of various noises (rain, wind, competing biophony) on the performance of the optimised model. All these results are discussed in the Discussion section, which integrates the modelling of the active detection area and explores its variations according to background noise.

# 2   Materials and methods

## 2.1   The convolutional neural network

TABLE 1 – Neural network architecture.

| Layer | Composition (type, activation function, max pooling) | Convolution parameters (Kernel size, Stride, Padding) |
|---|---|---|
| Conv1 | Conv2d (1 → 16) + ReLU + MaxPool2d (2, 2) | K=(3, 3), S=(2, 2), P=(2, 2) |
| Conv2 | Conv2d (16 → 32) + ReLU + MaxPool2d (2, 2) | K=(3, 3), S=(1, 1), P=(1, 1) |
| Conv3 | Conv2d (32 → 64) + ReLU + MaxPool2d (2, 2) | K=(3, 3), S=(1, 1), P=(1, 1) |
| Conv4 | Conv2d (64 → 32) + ReLU + MaxPool2d (2, 2) | K=(3, 3), S=(1, 1), P=(1, 1) |
| Lin1 | Linear (224 → 32) + ReLU | - |
| Lin2 (Decision) | Linear (32 → 1) + Sigmoid | - |

The architecture and input data of the convolutional neural network used for Rock Ptarmigan call recognition are directly inspired by a network developed as part of the COAT[6] project for monitoring populations of Rock Ptarmigan and Willow Ptarmigan (*Lagopus lagopus*) in Norway[1]. The network, of simple design, contains only 45,881 parameters, making it easily trainable and low in computational and energy consumption. It consists of four convolutional layers with ReLU activation functions, each followed by a max pooling layer (Table 1). Two successive fully connected layers complete the network. The probability that the sound sample provided as input is a Rock Ptarmigan call is obtained by a sigmoid function connected to the last layer. The final decision of the network is based on this probability: if it is greater than 0.5, the sample is classified as containing a Rock Ptarmigan call, at least in part; otherwise, it is classified as not containing one.

---

[1] The two networks developed as part of the COAT project achieve recall and precision values of 99.6% and 95.2% respectively for the network dedicated to the detection of Rock Ptarmigan calls, and 99.1% and 95.3% for the one dedicated to the recognition of Willow Ptarmigan calls.

The individual inputs to the network are preprocessed from 0.66-second sound samples. The raw signal is first downsampled to 8 kHz, then normalised between -1 and 1, and transformed into a spectrogram (Blackman-Harris windowing, step size of 0.016 s, 85% overlap, Slaney's[7] Mel scale across 40 bands from 400 Hz to 4 kHz, with levels in dB referenced to 1).

## 2.2 Training process

The data used for validation (*validLago*), testing (*testLago*) and partly for training (*trainLago*) come from recordings made in areas where Rock Ptarmigan is present in Norway (Svalbard and Varanger Peninsula) and France (Alps and Pyrenees). Wildlife Acoustics SM4 recorders[8] were used to collect these data. They were then manually annotated by several specialists for each recording session (or complete audio file), ensuring precise and consistent labelling. To avoid bias, the different recording sessions are not mixed between test, validation and training samples. To increase the robustness of the network, the training dataset was supplemented with external data (*trainExt*). Some of these data come from Rock Ptarmigan recordings downloaded from the participatory website Xeno-Canto[9], offering additional variety and quantity of target calls. These recordings were manually re-annotated to ensure their quality. The other part comes from recordings made as part of other ecoacoustic studies in areas without Rock Ptarmigan (French Guiana, Kuwait and France) or from the open access database FSD50K[10], containing only negative samples. Details of the datasets used are presented in Table 2.

TABLE 2 – Datasets composition.

| Dataset | Nb positives | Nb negatives | Nb files |
|---|---|---|---|
| *trainLago* | 21433 | 143802 | 292 |
| *trainExt* | 1791 | 21526 | 5522 |
| *trainAugm* | 4644 | - | - |
| *validLago* | 5981 | 27635 | 70 |
| *testLago* | 8500 | 26320 | 70 |

For network training, regularisation dropout layers are inserted after each convolutional layer with a dropout rate of 20%, and after the first fully connected layer with a dropout rate of 50%. The cost function used is binary cross-entropy. Optimisation is performed using the Adam algorithm with a learning rate of $10^{-5}$. Training is carried out over 1,000 epochs with batches of 32 signals (batch size). An initial training (conf0) is performed without adding augmented data, using only the training samples described previously (Table 3). This model serves as a reference to study the effect of adding augmented data. For the training of the three other models (Table 3), an augmented dataset (*trainAugm*) is added. This set contains only positive signals, representing 20% of the positive signals from the *trainLago* and *trainExt*

---

[2] 12 order, zero phase, bandpass Butterworth filter.

---

sets. These samples are constructed by superimposing two signals: one containing Rock Ptarmigan call, the other containing none. Signals without call were randomly selected from negative samples. Signals with call were sorted from positive signals with the instruction for the annotator to be clearly audible and have minimal background noise. To ensure the emergence of Rock Ptarmigan calls and because these calls are pulsed with inter-pulse durations greater than 0.01 s (Figure 1), only signals with a difference of at least 20 dB between the fractile levels $L_5$ and $L_{95}$ were retained. These fractile levels were calculated from 0.01 s windows in the frequency band from 400 Hz to 4 kHz[2]. The relative levels of signals containing call and those containing only noise are adjusted according to *snr* ranges varying according to the training configurations. The selected *snr* ranges (conf1 to conf3) each cover 10 dB and were defined according to the performance $p_0(snr)$ of the conf0 model (Figure 2): $p_0(snr) > 0.99$ for conf1, $0.88 > p_0(snr) > 0.12$ for conf2, and $p_0(snr) < 0.01$ for conf3.

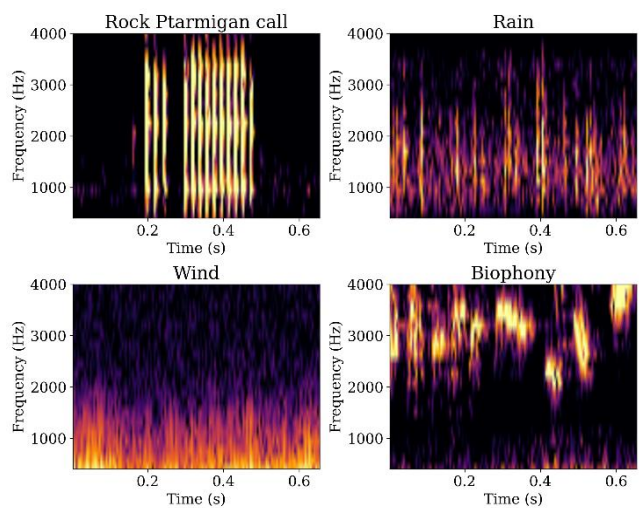

Figure 1 – Spectrograms of a male rock ptarmigan's call, rain noise, wind noise and biophonic noise, based on recordings made with SM4 (Wildlife) devices in areas where rock ptarmigan are present.

TABLE 3 – Training process configurations.

| Conf. | Training dataset | *snr* for trainAugm (dB) (corresponding $p(snr)$ for conf0) |
|---|---|---|
| 0 | trainLago + trainExt | - |
| 1 | trainLago + trainExt + trainAugm | [0 ; 10]   (>0.99) |
| 2 | trainLago + trainExt + trainAugm | [-16 ; -6]   (0.12 ;0.88) |
| 3 | trainLago + trainExt + trainAugm | [-36 ; -26]   (<0.01) |

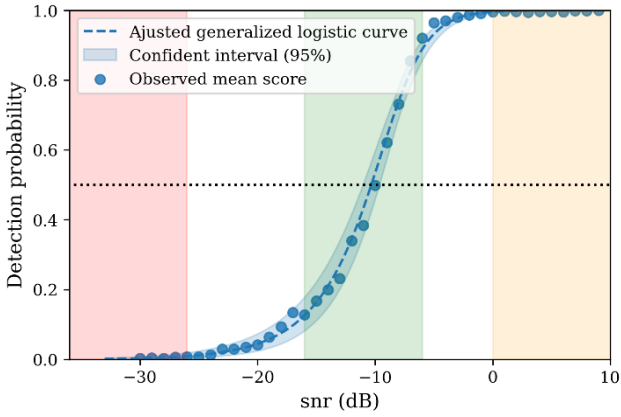

Figure 2 – Mean scores and generalized logistic function fit of acoustic detection performance for the conf0 training configuration (without augmented data).

🟧 *snr* range for conf1 augmented data
🟩 *snr* range for conf2 augmented data
🟥 *snr* range for conf3 augmented data

## 2.3 Detection probability as a function of signal-to-noise ratio and noise type

For each model trained according to a given configuration, the detection probability *p(snr)* is determined from artificial data constructed following the pattern of the *trainAugm* dataset (containing only positive data to which background noise is added). However, the initial data come from *validLago* and *testLago*, which were not used during training to avoid positive bias. The *snr* values considered range from -30 dB to 10 dB in steps of 1 dB. For each of these 40 values, 1,000 signals are generated. The corresponding *p(snr)* value is the average of the 1,000 inferences of the model considered at each *snr* value. The *p(snr)* curves present an "S" shape, often estimated by a sigmoid function. However, the sigmoid is mathematically symmetrical and does not account for potential differences in inflection for values close to 1 and 0. It is therefore preferable to use an approximation by a generalised logistic function with three parameters $x_0$, $k$ and $v$, where $x_0$ reflects the position of the inflection point on the *snr* values, $k$ the overall inflection and $v$ the inflection asymmetry:

$$p(snr) = \frac{1}{(1+e^{-k(snr-x_0)})^{\frac{1}{v}}} \quad (2)$$

The search for the parameters of this function that best estimate the measured *p(snr)* values was carried out by minimising the log-likelihood, and the uncertainties on these parameters were evaluated by *Bootstrap*[8] with 1,000 repetitions. The overall performance can be measured by $x_0$ or the *snr* value at 50% detection $snr_{50}$. The rate of performance variation can be measured by the inflection value at 50% detection $infl_{50}$ or more directly by the width of the inflection zone between 10% and 90% detections $\Delta snr_{5\text{-}95}$. The lower the $snr_{50}$, the more efficient the system is against noise. The lower the $\Delta snr_{5\text{-}95}$, the faster the transition from poor to good performance with the *snr*.

To study the performance *p(snr)* of the system as a function of noise, we considered four evaluation datasets, differentiated by the type of background noise. The first dataset contains signals with all types of noise, possibly including saturated signals. As the sampling is done equiprobably, the occurrences of the different types of noise are the same as in the test and validation samples. The other three datasets contain non-saturated signals: light rain noise, stable and broadband noise (mainly due to wind), or biophony noise (mainly due to avifauna). A pre-selection of these signals was carried out using inferences from the *YAMNet* sound event recognition network (for classes equivalent to rain, wind and biophony), followed by manual verification.

## 2.4 Detection area model

The level of *snr* is, by definition, obtained by subtracting the level of a Rock Ptarmigan call from the level of background noise (see Eq. (1) with $L_u = L_{Lago}$). If it is possible to estimate this *snr* in real conditions, it then becomes easy to use the *p(snr)* curve to determine the probability of detecting a call. Background noise pressure levels were recorded at the recorder, with measured sensitivities of 0.2 µ/Pa for recorders installed in France and 12.6 µ/Pa for those installed in Norway. The pressure level of the Rock Ptarmigan call at 1 metre is estimated at 85 ± 2 dB[10]. If we make the approximation that the Rock Ptarmigan is an omnidirectional source and that acoustic propagation is linear, without atmospheric losses or reflections, and on a flat ground, the pressure level of the call at a distance *r* is influenced only by the geometric dispersion of the sound wave:

$$L_{Lago}(r) = L_{Lago}(r = 1m) - 20\log_{10}(r) \quad (3)$$

In this framework, by inverting relation (1) and for a given noise type *n*, it is possible to estimate the distance *r* for a given detection probability value *p*:

$$r(p, L_n) = 10^{\frac{L_{Lago}(r=1m)}{20}} \cdot 10^{\frac{-L_n}{20}} \cdot 10^{\frac{-x_0}{20}} \cdot 10^{\frac{\frac{1}{k}\ln(p^{-v}-1)}{20}} \quad (4)$$

Assuming that the receiver is also omnidirectional, then the detection area $A(p, L_n)$ is given by the area of the disc of radius *r* :

$$A(p, L_n) = \pi \cdot r^2(p, L_n) \quad (5)$$

$$A(p, L_n) = \pi \cdot 10^{\frac{L_{Lago}(r=1m)}{10}} \cdot 10^{\frac{-x_0}{10}} \cdot 10^{\frac{-L_n}{10}} \cdot 10^{\frac{\frac{1}{k}\ln(p^{-v}-1)}{10}} \quad (6)$$

## 3 Results

### 3.1 Performances and adapted augmented data

To quantify the performance of the neural network, particularly regarding the training configuration, we selected classical machine learning metrics: the value of the cost function (or loss), weighted accuracy (to account for the imbalance in the number of positive and negative samples in the dataset), precision, recall, and the F1 measure (geometric mean of precision and recall). The values of the loss function and weighted accuracy as a function of the training epoch (Figure 3) improve rapidly over the first 200 epochs, then stabilise at values of good overall performance. These curves and all the values of machine learning metrics for the final models, after the thousandth epoch (Table 4), show a weak effect of the training configuration on the validation and test samples, whereas, except for precision, this effect is more pronounced on the training sample. It should be noted that only for conf2, the learning curves and precision, recall and F1 measure tend to converge between the training sample and those of validation and testing.

The acoustic detection metrics (Table 5), derived from $p(snr)$ (Figure 4), show homogeneous performance for conf0, conf1 and conf3. Conf2 produces a shift of the $p(snr)$ curve of about 1 dB towards negative values ($snr_{50}$ or $x_0$) compared to the other three configurations, indicating better performance.

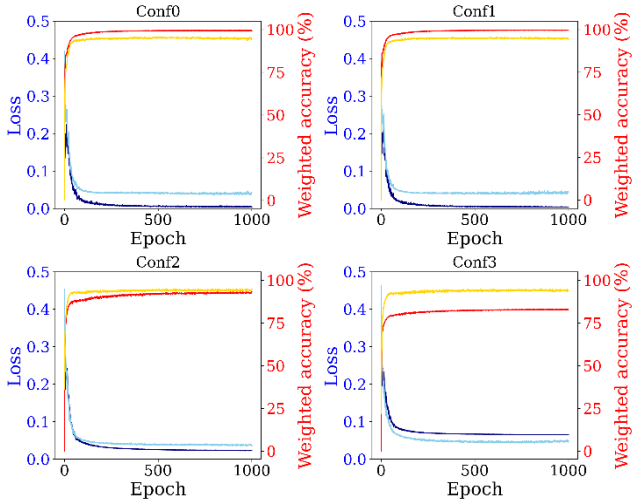

Figure 3 – Learning curves of the four CNN models, differentiated by the training dataset (with or without different augmented dataset). ▬: training loss, ▬: validation loss, ▬: training weighted accuracy, ▬: validation weighted accuracy.

TABLE 4 – Machine learning performance metrics.

| Conf. | Sample | Loss | W. acc | Prec. | Recall | F1 |
|---|---|---|---|---|---|---|
| 0 | Training | 0.0041 | 99.2 | 99.5 | 99.3 | 99.4 |
| 0 | Validation | 0.0435 | 94.4 | 99.7 | 94.5 | 97.0 |
| 0 | Test | 0.0368 | 96.5 | 99.7 | 96.6 | 98.1 |
| 1 | Training | 0.0038 | 99.6 | 99.6 | 99.6 | 99.6 |
| 1 | Validation | 0.0425 | 94.7 | 99.5 | 94.8 | 97.1 |
| 1 | Test | 0.0439 | 96.4 | 99.6 | 96.6 | 98.0 |
| 2 | Training | 0.0229 | 93.1 | 99.5 | 93.1 | 96.2 |
| 2 | Validation | 0.0358 | 94.8 | 99.6 | 94.9 | 97.2 |
| 2 | Test | 0.0372 | 96.6 | 99.7 | 96.6 | 98.1 |
| 3 | Training | 0.0651 | 82.9 | 99.9 | 82.9 | 90.6 |
| 3 | Validation | 0.0482 | 94.0 | 99.8 | 94.0 | 96.8 |
| 3 | Test | 0.0483 | 95.8 | 99.9 | 95.7 | 97.7 |

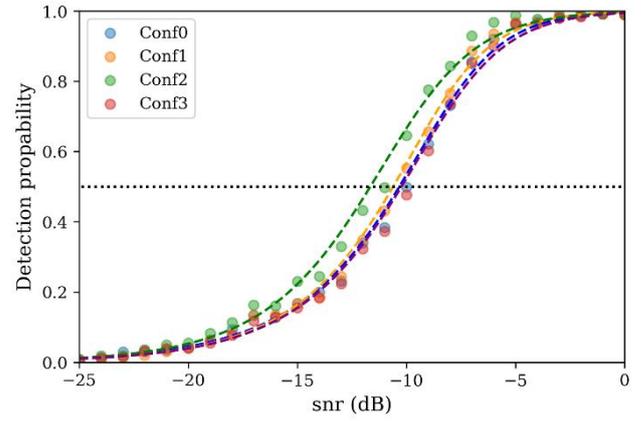

Figure 4 – Average measurements (●) and their adjustments by a generalized logistic function (--) of the acoustic performance in detection $p(snr)$ according to the training configuration.

TABLE 5 – Acoustic performance measurements in detection (without distinction on the type of noise). The configuration with the best performance is greyed out.
$snr_{50}$: the $snr$ value at 50% detection, related to the global performance.
$infl_{50}$: the inflection value at 50% detection (the derivative).
$\Delta snr_{5-95}$: the width of the inflection zone between 10% and 90% detections.
$x_0$, $k$ and $v$ are the parameters of the generalised logistic function approximating $p(snr)$ (see Eq. (2)), respectively related to its global position on the $snr$ axis, its inflection and its asymmetry.

| Conf. | $snr_{50}$ (dB) | $infl_{50}$ (dB$^{-1}$) | $\Delta snr_{5-95}$ (dB) | $x_0$ (dB) | $k$ (dB$^{-1}$) | $v$ |
|---|---|---|---|---|---|---|
| 0 | -10.3 | 0.10 | [-19.6 ; -4.4] | -8.1 | 0.58 | 2.2 |
| 1 | -10.6 | 0.11 | [-19.3 ; -4.8] | -8.6 | 0.58 | 2.0 |
| 2 | -11.7 | 0.10 | [-20.2 ; -5.1] | -10.1 | 0.48 | 1.6 |
| 3 | -10.2 | 0.10 | [-19.2 ; -4.1] | -8.3 | 0.52 | 1.9 |

### 3.2 Noise type effect on detection probability

Considering only the model from conf2, we observe a significant influence of noise type on acoustic detection performance (Table 5 and Figure 5). When the background noise is solely produced by rain, the neural network performs worse than in the case of average background noise, with a difference of about 3 dB. On the other hand, for background noise produced by light wind or consisting of biophony, the system performs better than average, with respective gains of about 4 dB and 9 dB.

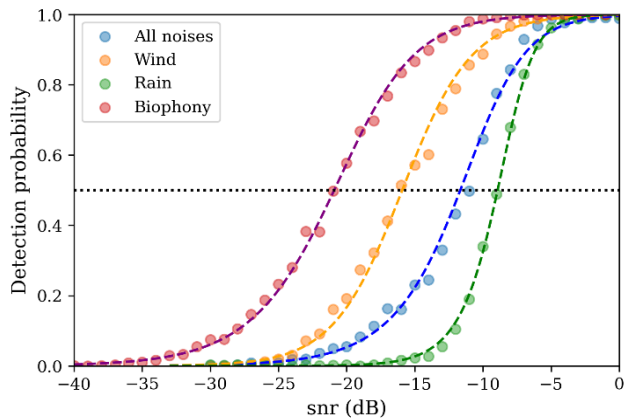

Figure 5 – Mean measurements (●) and their adjustments by a generalised logistic function (--) of acoustic detection performance *p(snr)* according to the type of background noise for conf2.

TABLE 6 – Acoustic detection performance measurements for model 2 (conf2) according to noise type.
$snr_{50}$: the *snr* value at 50% detection, related to the global performance.
$infl_{50}$: the inflection value at 50% detection (the derivative).
$\Delta snr_{5-95}$: the width of the inflection zone between 10% and 90% detections.
$x_0$, $k$ and $v$ are the parameters of the generalised logistic function approximating *p(snr)* (see Eq. (2)), respectively related to its global position on the *snr* axis, its inflection and its asymmetry.

| Bruit | $snr_{50}$ (dB) | $infl_{50}$ (dB$^{-1}$) | $\Delta snr5\text{-}95$ (dB) | $x_0$ (dB) | $k$ (dB$^{-1}$) | $v$ |
|---|---|---|---|---|---|---|
| All type | -11.7 | 0.10 | [-20.2 ; -5.1] | -10.1 | 0.48 | 1.6 |
| Wind | -15.9 | 0.10 | [-23.4 ; -8.5] | -15.9 | 0.40 | 1 |
| Rain | -8.9 | 0.18 | [-13.8 ; -5.3] | -8.0 | 0.87 | 1.7 |
| Biophony | -20.9 | 0.08 | [-30.7;-12.6] | -19.7 | 0.37 | 1.4 |

# 4 Discussion

## 4.1 Neural network performance and data augmentation effect

Comparing machine learning metrics between training samples and test and validation samples provides valuable insights into the network's learning process. On the test and validation samples, these metrics change little according to the training configuration and are substantially identical, indicating a certain independence from the training configuration and good homogeneity between these two samples. The loss, recall, weighted accuracy and F1 measure on the training sample show, conversely, a divergence from the test or validation samples, a difference that evolves according to the augmented data sample added to the training. For the training sample comparing conf0 and conf1, these metrics are identical or better than for the test and validation samples. At first glance, this could reflect slight overfitting, but it should be remembered that the training sample has a different composition from the test/validation samples; it contains additional external data *(trainExt)*. For conf2, these metrics and learning curves tend to converge between the different samples, producing an apparent regularisation effect on the network. Conf3 sees these metrics on the training sample become worse than on the test/validation samples, reflecting slight underfitting probably due to the network's difficulty in learning to detect target calls in excessively deleterious background noise conditions.

Machine learning performance metrics on test and validation samples are substantially identical across training configurations, whereas acoustic detection evaluation *p(snr)*, made from these same samples, indicates better performance for conf2. Thus, machine learning performance metrics cannot be directly linked to acoustic detection performance. These two metrics are therefore complementary, providing distinct information.

The addition of augmented data by constructing noisy positive signals to the training sample significantly influences the acoustic performance *p(snr)* of the model only when the *snr* values of these augmented data are in the transition zone (conf2) of $p_0(snr)$ of the initial model without augmented data (conf0). Adding augmented data for higher *snr* values does not improve performance, as the initial model is already very efficient (conf1). For *snr* values that are too low, for which the initial model hardly works (conf3), acoustic detection performance is not improved either. Thus, performance improvement is not induced by adding data to the training sample (conf0 compared to conf1). When generating augmented data, it is by targeting *snr* values in the inflection zone of $p_0(snr)$ that network performance can be improved.

## 4.2 Detection area model

The study of equation (6) shows theoretically that for a detection probability limit *p*, a decrease in background noise level or an improvement in acoustic detection performance ($x_0$) of just 1 dB increases the detection area of the recorder by 26%. Conversely, an increase in background noise or a decrease in acoustic performance of 1 dB results in a loss of 21% of this area. Thus, conf2, even though it shows an improvement in $x_0$ of only 1.5 to 2 dB compared to other configurations, significantly increases the detection area by 41% to 58%, thereby influencing the estimation of the surface density of calls.

Figure 6 shows that the detection area is very sensitive to noise and can vary by several hundred square metres, depending on the type and level of background noise. The coverage area of the recorder, and therefore also the estimation of the surface density of calls, are highly dependent on the sound environment.

In real conditions, the shape and size of detection areas can deviate widely from a circular area [11]. More advanced modelling of propagative phenomena including terrain conditions (meteorological, nature and topography of terrain, etc.) [10] combined with knowledge of *p(snr)* would provide a more refined estimation of the detection area. Strictly speaking, the modelling proposed here, based solely on geometric attenuation, can only be considered for estimating orders of magnitude on these areas and their variations.

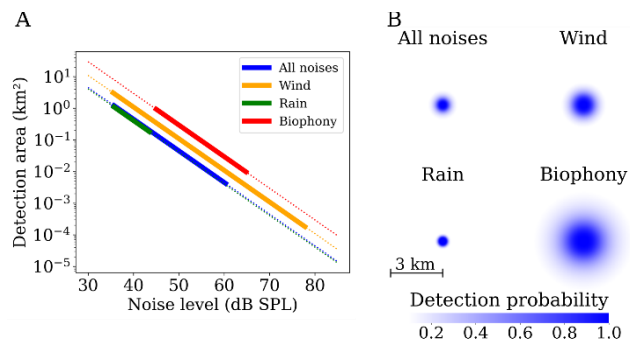

Figure 6 – Detection area according to noise type for the model trained according to conf2. A) for *p(snr)*=0.9 and as a function of background noise level (in thick lines: the 5%-95% inter-percentile of measured SPL levels). B) for a background noise level of 45 dB.

## 5   Conclusion

We estimated the acoustic detection performance *p(snr)* of a neural network dedicated to detecting the call of a target species by creating an artificial dataset by superimposing extracts of the target call and noise signals while controlling the signal-to-noise ratio.

We showed that knowledge of *p(snr)* provides additional information on network performance compared to using only classical machine learning metrics. Furthermore, we demonstrated that this knowledge allows optimising the network's performance on this same metric, thanks to the addition of augmented data whose snr has been targeted at the transition values of *p(snr)*. We also observed that *p(snr)* can be significantly influenced by variations in the sound environment.

Based on simplified modelling, we also proposed a simulation of the detection area. We showed that this area can be very sensitive to the acoustic performance *p(snr)* and, through it, also to the type and levels of background noise

## Acknowledgements

This research was funded as part of the COAT (*Climate-ecological Observatory for Arctic Tundra*) program. We would especially like to thank John-André Henden, Nigel Yoccoz and Eva Fuglei for the recordings and their patient manual annotations. We would also like to thank Jérôme Sueur for providing the external recordings and Hugo Robert and Yoann Meignant for their annotations. Finally, we would like to thank Arthur Guibard and Léo Papet for estimating the sensitivity of the SM4 recorders.